\documentclass[10pt,twocolumn,letterpaper]{article}

\usepackage{wacv}
\usepackage{times}
\usepackage{epsfig}
\usepackage{graphicx}
\usepackage{amsmath}
\usepackage{amssymb}
\usepackage{subcaption}
\usepackage{float}
\usepackage{multirow}
\usepackage{placeins}
\usepackage{booktabs}
\usepackage[flushleft]{threeparttable}
\newcommand{\tmop}[1]{\ensuremath{\operatorname{#1}}}

\def\wacvPaperID{166} 

\wacvfinalcopy 

\ifwacvfinal
\def\assignedStartPage{1} 
\fi

\ifwacvfinal
\usepackage[breaklinks=true,bookmarks=false]{hyperref}
\else
\usepackage[pagebackref=true,breaklinks=true,colorlinks,bookmarks=false]{hyperref}
\fi
\usepackage{cleveref}
\ifwacvfinal
\else
\fi

\begin{document}

\title{Spatial Context-Aware Self-Attention Model For Multi-Organ Segmentation}

\author{Hao Tang, Xingwei Liu, Kun Han, Shanlin Sun, Narisu Bai, \\ Xuming Chen, Huang Qian, Yong Liu, Xiaohui Xie\thanks{Corresponding author}\\
University of California Irvine and 
School of Medicine Shanghai Jiao Tong University 
}





\maketitle

\begin{abstract}
Multi-organ segmentation is one of most successful applications of deep learning in medical image analysis. Deep convolutional neural nets (CNNs) have shown great promise in achieving clinically applicable image segmentation performance on CT or MRI images. State-of-the-art CNN segmentation models apply either 2D or 3D convolutions on input images, with pros and cons associated with each method: 2D convolution is fast, less memory-intensive but inadequate for extracting 3D contextual information from volumetric images, while the opposite is true for 3D convolution. To fit a 3D CNN model on CT or MRI images on commodity GPUs, one usually has to either downsample input images or use cropped local regions as inputs, which limits the utility of 3D models for multi-organ segmentation. In this work, we propose a new framework for combining 3D and 2D models, in which the segmentation is realized through high-resolution 2D convolutions, but guided by spatial contextual information extracted from a low-resolution 3D model. We implement a self-attention mechanism to control which 3D features should be used to guide 2D segmentation.  Our model is light on memory usage but fully equipped to take 3D contextual information into account. Experiments on multiple organ segmentation datasets demonstrate that by taking advantage of both 2D and 3D models, our method is consistently outperforms existing 2D and 3D models in organ segmentation accuracy, while being able to directly take raw whole-volume image data as inputs.


\end{abstract}
\section{Introduction}
Segmentation of organs or lesions from CT images has great clinical implications. It can be used in multiple clinical workflows, including diagnostic interventions, treatment planning and treatment delivery \cite{gibson2018automatic}. Organ segmentation is an importance procedure for computer-assisted diagnostic and biomarker measurement systems \cite{van2011computer}. Organ-at-risk (OAR) segmentation and tumor segmentation are also crucial to the planning of radiation therapy \cite{sykes2014reflections}. Moreover, the segmentation-based models of anatomical structures can support surgical planning and delivery \cite{howe1999robotics}.

Organ segmentation is typically done manually by experienced doctors. However, manually segmenting CT image by doctors is often time consuming, tedious and prune to human error, as a typical CT scan can contain up to hundreds of 2D slices. Computational tools that automatically segment organs from CT images can greatly alleviate the doctors' manual effort, given a certain amount of accuracy is achieved.

There is a vast volume of work on organ segmentation using CT or magnetic resonance (MR) image. Traditional segmentation methods are mostly atlas-based. These methods rely on a set of accurate image templates with manual segmentation, and then use image registration to align the new image to the templates. Because of the reliance on the pre-computed templates, these methods may not adequately account for the anatomical variance due to variations in organ shapes, removal of tissues, growth of tumor and differences in image acquisition \cite{wu2019aar}. Also, registration is computationally intensive and may take up to hours to complete \cite{daisne2013atlas,fortunati2013tissue,duc2015validation,hoogeman2008atlas,levendag2008atlas,qazi2011auto,sims2009pre,teguh2011clinical,thomson2014evaluation,walker2014prospective,voet2011does,isambert2008evaluation,fritscher2014automatic,commowick2008atlas,verhaart2014relevance,wachinger2015contour,fortunati2015automatic,zhang2007automatic}. 

Deep learning-based methods provide an alternative solutions with substantial accuracy improvement and speed-up. With recent advances in deep learning especially deep convolutional neural network, automatic segmentation using computer algorithm has shown great promise in achieving near human performance \cite{bai2017human,zhao20183d,liao2019evaluate,tang2019clinically}, and various applications have been deployed in clinical practice. 

Fully convolutional network \cite{long2015fully} and U-Net \cite{ronneberger2015u} are two of the most widely used deep learning-based segmentation algorithms for this purpose. 
Many its variants have been proposed in recent years, including V-Net \cite{milletari2016v} and Attention U-Net \cite{oktay2018attention}. These methods can use either 2D or 3D convolutions as its basic building component. 2D methods usually operate on a slice by slice basis, while 3D methods often operate on a 3D block or a stack of multiple 2D slices \cite{liu20183d, zhou2016three}. The whole volume prediction can be obtained by predicting each slice or block using a sliding window. Additionally, some may stack multiple 2D slices in the input channel and use 2D convolution as a way to include some 3D features, and this is often referred as 2.5D model.

However, a CT image is inherently 3D. Cutting the images into slices or blocks often ignores the rich information and relation within the whole image volume. A big challenge in developing algorithm for consuming the whole image volume is the GPU memory limitation. Simply storing the tensors of the image features
would require huge amount of GPU memory. One way is to adopt a coarse-to-fine strategy \cite{zhu20183d,yu2018recurrent,wang2019abdominal,zhao2019fully,wang2020fully,ma2018novel,xia2018bridging,zhu20183d,yu2018recurrent,zhou2017fixed,cai2017improving,roth2018spatial}, where in the first stage the organs of interest are roughly located, and in the second stage the segmentation masks are further refined by using a smaller input based on the localization. This usually requires training multiple CNNs for different stages and organs. Recently, several methods have been proposed to use the whole CT image volume as input, and achieve state-of-the-art accuracy and inference speed \cite{tang2019clinically,zhu2019anatomynet,zhu20183d,tang2019nodulenet,guo2020organ}. Despite their successes, there exists several disadvantages. First, to reduce the GPU memory consumption, these methods usually directly downsample the input in the very first convolution layer, which may lead to loss of local features. Moreover, they require carefully tailored image input size in order to fit the whole-volume image. However, they will still face GPU memory limitation if the image resolution becomes higher, because the memory requirement grows quickly with the size of the image volume. This makes previous whole-volume algorithm less effective when adapted to new dataset.
Second, some of them make strong assumption on the organs/region they segment, thus lacking the ability to generalize well to other parts of the CT image. 

We seek to incorporate 3D whole volume information into 2D model in a scalable way. We hypothesize that the benefits of using 3D convolution on the whole-volume image may come from its capability of modeling the shapes and relationships of the 3D anatomical structures. However, to model such shapes and relationships, we do not have to use very high-resolution image. 3D convolution on downsampled image volume may suffice to extract such information and can save a lot of computation and GPU memory. We can use 2D convolution on the original image slice to compensate for the loss of resolution. To fuse both 3D context features and 2D features, we implement a new module called multi-slice feature aggregation based on self-attention \cite{vaswani2017attention}, 
which treats the 2D feature map as query and 3D context map as key, and uses self-attention to aggregate the rich 3D context information.


In this paper, we propose a new deep learning framework named \textbf{S}patial \textbf{C}ontext-\textbf{A}ware Self-\textbf{A}ttention Model (SCAA). Our main contributions are: \romannumeral1) a new framework for combining 3D and 2D models that takes the whole-volume CT image as input; \romannumeral2) a self-attention mechanism to filter and aggregate 3D context features from the whole volume image to guide 2D segmentation \romannumeral3) the proposed method can scale to larger input volume without concerning the GPU memory limitation that common 3D methods face. Experiment results on a head and neck (HaN) dataset of 9 organs and an abdomen dataset of 11 organs show the proposed model consistently outperforms state-of-the-art methods in terms of organ segmentation accuracy, while being able to take the whole-volume CT image as input.


\section{Method}
\begin{figure*}
\begin{center}
 \includegraphics[width=0.9\textwidth]{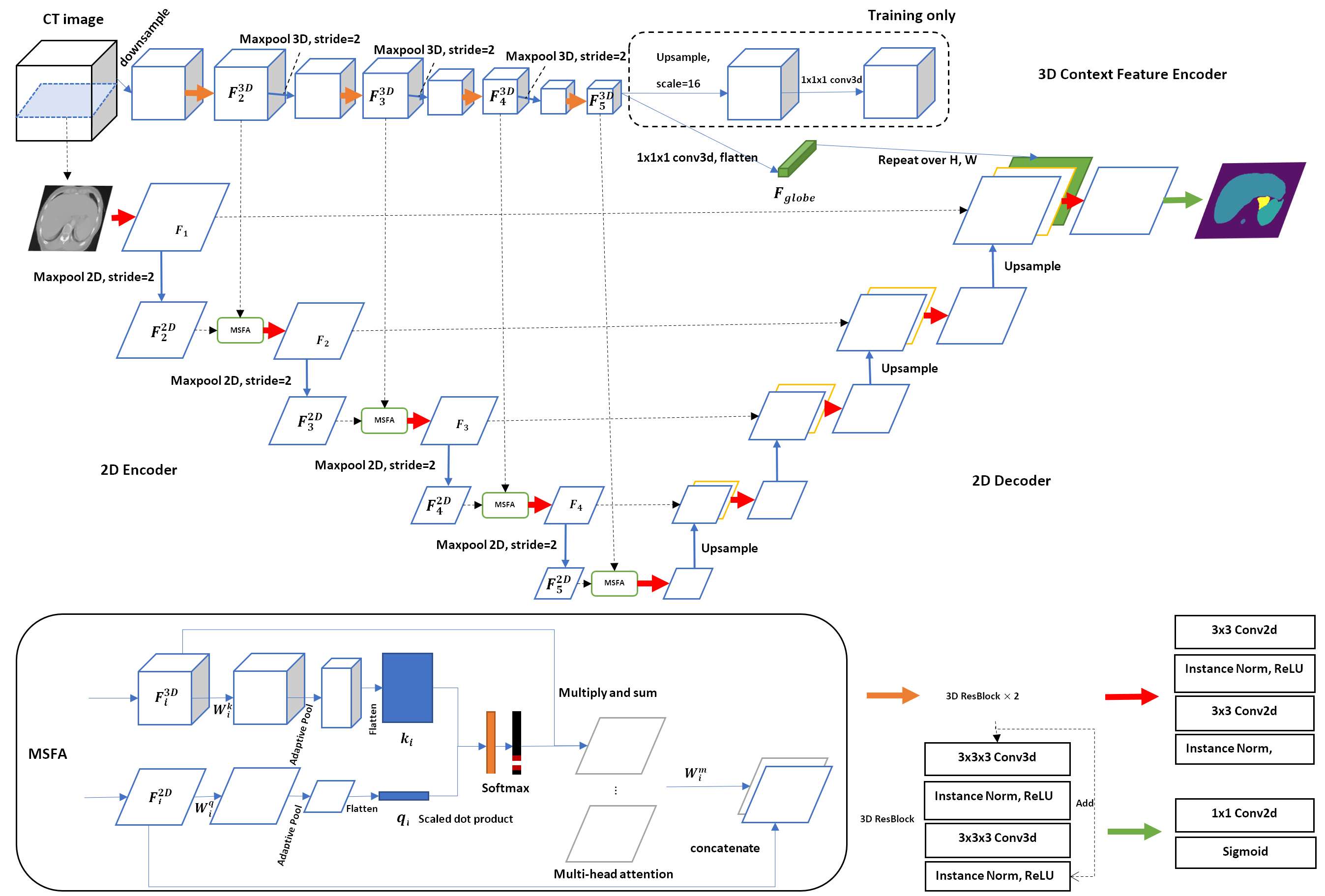}
\end{center}
\caption{\textbf{Overview of spatial context-aware self-attention model (SCAA)}. SCAA consists of a 3D context feature encoder, a 2D encoder, a 2D decoder and a multi-slice feature aggregation (MSFA) module. SCAA starts with extracting 3D features from the downsampled CT image using 3D convolutions. Then the 2D encoder extracts 2D features and uses MSFA module to fuse 2D and 3D features hierarchically. Lastly, the 2D decoder decodes the fused 2D and 3D features and outputs 2D segmentation masks of each organ. The numbers of feature channels in $F_i$ are {64, 96, 128, 192, 256} for $i=1,2,3,4,5$ respectively. The number of feature channels in $F_i^{2D}$ is {96, 128, 192, 256} for $i=2,3,4,5$ respectively. The numbers of feature channels in $F_i^{3D}$ are {24, 32, 64, 64} for $i=2,3,4,5$ respectively. 
$W_i^k$ is implemented by using $1\times 1\times 1$ 3D convolution with 2, 2, 4, and 4 feature channels for $i=2,3,4,5$ respectively. $W_i^q$ is implemented by using $1\times 1$ 2D convolution with 2, 2, 4, and 4 feature channels for $i=2,3,4,5$ respectively. The $xy$ spatial resolution of the output of the adaptive pooling are $16\times16$, $8\times8$, $4\times4$, $4\times 4$ for scale $i=2,3,4,5$ respectively. The number of attention heads for scale $i=2,3,4,5$ is 2, 2, 4, and 4 respectively.
}
\label{fig:model}
\end{figure*}

\Cref{fig:model} shows the details of the proposed method. The proposed model consists of four parts: a 3D context feature encoder $f^{3D}$, a 2D feature encoder $f^{Enc}$, a multi-slice feature aggregation (MSFA) module $f^{MSFA}$, and at last a 2D decoder $f^{Dec}$. The input to the model $f: x\rightarrow f^{Dec}(f^{Enc}(f^{3D}(x)))$ is the whole CT image volume $\mathbf{I} \in \mathbb{R}^{D\times H\times W}$, and the outputs are $D$ 2D segmentation masks for $C$ classes $\mathbf{m} \in \mathbb{R}^{D\times H\times W \times C}$. $D, H, W$ are the depth, height and width of the image volume.

\subsection{3D context feature encoder and 2D encoder}
$f^{3D}$ first downsamples the input 3D volume $\mathbf{I}$ to $\mathbf{I'} \in \mathbb{R}^{D^{3D}\times H^{3D}\times W^{3D}}$. $D^{3D}, H^{3D}, W^{3D}$ are the depth, height, and width of the downsampled 3D volume. We use a downsample factor of two in this work. Note we can also downsample the volume to other resolutions, e.g. isotropic 4mm resolution. It then applies 3D convolution blocks three times, where each convolution block consists of two residual blocks followed by one 2 $\times$ 2 $\times$ 2 max pooling, aiming at extracting 3D context features in the whole CT image. 


The output of $f^{3D}$ are four feature maps at different scales, denoted as $F^{3D}_i \in \mathbb{R}^{C^{3D}_i \times D^{3D}_i \times H^{3D}_i \times W^{3D}_i}$, where $i={2, 3, 4, 5}$. This means the feature map $F^{3D}_i$ is downsampled by a factor of $2^i$ compared to the original image. $D_i^{3D}$, $H^{3D}_i$ and $W^{3D}_i$ are the depth, height and width of the feature map at scale $i$, the values of which depend on the size of input image. $C^{3D}_i$ equals 24, 32, 64 and 64 for $i=2,3,4,5$ respectively. After $F^{3D}_5$, we flatten the channel, depth, height and width dimension into a vector and regard it as a global descriptor $F_{globe}$ for the 3D volume.

$f^{Enc}$ is similar to U-Net encoder. It consumes one axial slice of the CT image $\mathbf{S} \in \mathbb{R}^{H\times W}$ and applies 2D convolution blocks five times, where each block consists of two convolutions followed by instance normalization and ReLU activation, and a max pooling at the end. The 2D feature encoder outputs five sets of feature maps at different scales, denoted as $F^{2D}_i \in \mathbb{R}^{C^{3D}_i \times H^{2D}_i \times W^{2D}_i}$, where $i={1, 2, 3, 4, 5}$. $C^{2D}_i$ equals 64, 96, 128, 192 and 256 for $i=2,3,4,5$ respectively.

\subsection{Multi-scale feature aggregation}
Inspired by Transformer \cite{vaswani2017attention,ramachandran2019stand}, we implement a self-attention mechanism to filter and extract useful 3D context features from our 3D feature maps $F^{3D}_i$, and we name this module as multi-scale feature aggregation (MSFA). We regard 3D features as values, and generate queries from the 2D features and keys from the 3D features. Based on feature similarities of current 2D features and all slices in the 3D feature map (along $z$ dimension), the MSFA will generate an attention vector $\mathbf{a_i} \in \mathbb{R}^{D^{3D}_i}$ the same length of the depth of the 3D feature map. This attention then is applied to the 3D feature map to generate a 2D feature map that is considered as the aggregated 3D context features.

We start by mapping our 2D feature map $F^{2D}_i$ and 3D feature map $F^{3D}_i$ ($i={2, 3, 4, 5}$), to one query and $D_i^{3D}$ keys, of the same embedding space. We use a weight metric $W^q_i$ ($1\times 1$ 2D convolution) to generate our query $q \in \mathbb{R}^{C_{embed} \times H_i \times W_i}$. We use a weight metric $W^k_i$ ($1\times 1\times 1$ 3D convolution) to generate our keys $\{k_j\}$ of size $C_i^{embed} \times H^{3D}_i \times W^{3D}_i$, where $j={1,2,...,D^{3D}_i}$. An adaptive pooling operation is used to reduce the spatial resolution of the query and keys to $H_i' \times W_i'$, followed by a flatten operation to make them one dimensional. As a result, the embed dimension of the query and keys is now of size $C_i^{embed} \cdot H_i' \cdot W_i'$:

\begin{equation}
\label{equation:attention}
\begin{gathered}
(F^{3D'}_i)_{c'dhw} = (F^{3D}_i)_{c d h w} (W_i^k)_{c' c} \\
(F_i^{2D'})_{c'hw} = (F^{2D}_i)_{c h w} (W_i^q)_{c'c} \\
k_i = \tmop{Flatten} (\tmop{AdaptivePool}_{D_i^{3D}H'_iW'_i} (F^{3D'}_i)) \\
q_i = \tmop{Flatten} (\tmop{AdaptivePool}_{H'_iW'_i} (F_i^{2 D', }))
\end{gathered}
\end{equation}

$(\cdot)_{(\cdot)}$ is the Einstein summation convention.

A scaled dot product is used to compute the response of the 2D feature map with the 3D feature map $r_j=\frac{q\cdot k_j}{\sqrt{C_i^{embed}\cdot H'_i \cdot W'_i}}$. A softmax is followed to generate our attention $\mathbf{a_i}=\mbox{softmax}(\mathbf{r})$. We then multiply the attention $\mathbf{a_i}$ over the depth dimension of $F^{3D}_i$ and sum over the depth dimension to generate our aggregated context feature map $F^{agg}_i \in \mathbb{R}^{C^{3D}_i \times H_i \times W_i}$:
\begin{equation}
\label{equation:attention}
\begin{gathered}
(F^{agg}_i)_{chw} = (F^{3D}_i)_{cdhw} (\mathbf{a_i})_{d}
\end{gathered}
\end{equation}

A multi-head attention mechanism is also used. We generate $m_i$ such fused 2D feature map and then use a weight metric $W^m_i$ ($1\times 1$ 2D convolution) to aggregate multiple self-attention output. $m_i$ is 2, 2, 4 and 4 for $i=2,3,4,5$ respectively. This multi-head attention allows our model to focus on different parts of the 3D context volume to extract features required by different classes. Two 2D convolution blocks on the concatenated 2D feature map of $F^{2D}_i$ and $F^{agg}_i$ are used to better combine the 2D and 3D context features. $F_i^{}$ denotes our final 2D feature map for scale $i$. Note that $F_1$ is the same as $F_1^{2D}$.

\subsection{2D decoder}
$f^{Dec}$ is similar to the U-Net decoder. Starting from $F_5$, a 2D upsample is used first to increase the spatial resolution by 2. Then we concatenate the upsampled features with the corresponding encoder feature map and apply one 2D convolution block. The last upsampled feature map is of the same resolution as our input image. We concatenate our 3D global descriptor $F_{globe}$ to each pixel's feature vector and use a $1\times 1$ convolution to obtain the final axial segmentation mask for each class $\{\mathbf{m}^{2D}_c\}$, where $c\in \mathbb{Z}^{*}_{<C}$ and $C$ is the number of classes.

\subsection{Loss function and implementation details}
The loss function is defined as:
\begin{equation}
\label{equation:dice_loss}
\begin{gathered}
L^{2D} =  \sum_c^N{1 - \phi(\mathbf{m}^c, \mathbf{g}^c)}
\end{gathered}
\end{equation}

$\mathbf{g}$ is the ground truth segmentation for the axial slice. $\phi(\mathbf{m},\mathbf{g})$ computes a soft Dice score between the predicted mask $\mathbf{m}$ and the ground truth $\mathbf{g}$:
\begin{equation}
\begin{scriptsize}
\begin{gathered}
\phi(\mathbf{m}, \mathbf{g}) =  \frac{\sum_{i}^{N}\mathbf{m}_i \mathbf{g}_i}{\sum_{i}^{N}\mathbf{m}_i\mathbf{g}_i + \alpha \sum_{i}^{N}\mathbf{m}_i(1-\mathbf{g}_i) + \beta \sum_{i}^{N}(1-\mathbf{m}_i)\mathbf{g}_i + \epsilon}
\end{gathered}
\end{scriptsize}
\end{equation}
N is the number of total pixels in the batch. $\alpha$ and $\beta$ are two hyper parameters controlling the penalty for false positive and true negative respectively, and we set them to both 0.5. $\epsilon$ is used for numerical stability.

To facilitate the training of 3D context feature encoder, we add an auxiliary 3D segmentation loss. We first upsample $F^{3D}_5$ by a factor of 16, so it has the same spatial resolution as the downsampled 3D image volume. A $1\times 1\times 1$ 3D convolution is used to obtain the 3D segmentation mask $\{m^{3D}_c\}$. We use the same dice loss to get our 3D supervision loss $L^{3D}$. The final loss is then $L=L^{3D}+L^{2D}$.

We use one CT image for each batch during training. For each batch, we generate one 3D image volume and randomly sample 16 axial slices (batch size 16 for the 2D network). We only need to forward the 3D context encoder once per batch. We use Adam with initial learning rate $10^{-4}$ as optimizer for a total of 150 epochs. We applied elastic transformation and random jitter for data augmentation.

For testing, we only need to forward the 3D context feature encoder once, and we segment each 2D slice using the 2D decoder/encoder and MSFA.

\section{Experiments}
\subsection{Datasets}
Two datasets were used for evaluation: 
\romannumeral1) MICCAI 2015 head and neck (HaN)organ-at-risk (OAR) segmentation challenge dataset \cite{raudaschl2017evaluation}, containing a training of 33 CT images and a test of 15 CT images. The dataset contains manually labeled organ segmentation mask for 9 organs: brain stem, mandible, optic nerve left and right, optic chiasm, parotid left and right, submandibular gland (SMG) left and right; 
\romannumeral2) an in-house abdomen multi-organ segmentation dataset\footnote{Use of this dataset has been approved by an institutional review board (IRB).} (ABD-110) containing 110 contrast enhanced CT images and 11 organs (large bowel, duodenum, spinal cord, liver, spleen, small bowel, pancreas, left and right kidney, stomach and gallbladder). The 110 CT scans were collected from 110 patients who had radiotherapy during the past three years. The CT scans were manually delineated by one experienced doctor and then manifested by another. We use the official split of training set to train the model and test on the official test set on MICCAI 2015 challenge dataset, following the same protocol as previous work \cite{tang2019clinically,nikolov2018deep,raudaschl2017evaluation,guo2020organ}. All experiments on ABD-110 dataset was conducted using 4-fold cross validation. 

We report the segmentation performance using dice similarity coefficient (DSC) in percentage and 95\% hausdorff distance (HD) in mm following previous work \cite{raudaschl2017evaluation}.
DSC measures the overlap between the predicted mask $\mathbf{m}$ and ground truth mask $\mathbf{g}$:
\begin{equation}
\begin{gathered}
\mathbf{DSC} = \frac{2|\mathbf{m} \cup \mathbf{g}|}{|\mathbf{m} \cap \mathbf{g}|}
\end{gathered}
\end{equation}

\subsection{Ablation study on ABD-110}
\label{subsection:ablation}

\setlength{\tabcolsep}{2pt}
\begin{table*}
\centering
\begin{threeparttable}
\begin{scriptsize}
\begin{tabular}{l l}
\begin{tabular}{l l l l l l l}
\hline
Anatomy & U-Net & CA & C-CA & SCAA & SCAA$^*$ \\ \hline
Large Bowel  & $80.5 \pm 9.4$ & $79.6 \pm 10.2$ & $81.5 \pm 9.0$ & $81.5 \pm 10.0$ & $\textbf{82.5 $\pm$ 9.2}$ \\ 
Duodenum  & $63.4 \pm 18.6$ & $67.6 \pm 17.4$ & $69.9 \pm 17.2$ & $\textbf{71.4 $\pm$ 17.2}$ & $70.7 \pm 17.5$ \\ 
Spinal Cord  & $90.3 \pm 3.8$ & $90.4 \pm 3.8$ & $\textbf{91.0 $\pm$ 3.7}$ & $90.7 \pm 4.0$ & $90.8 \pm 3.5$ \\ 
Liver  & $95.5 \pm 1.9$ & $96.0 \pm 1.9$ & $96.2 \pm 1.4$ & $96.4 \pm 1.1$ & $\textbf{96.4 $\pm$ 1.2}$ \\ 
Spleen  & $94.6 \pm 3.1$ & $95.2 \pm 2.3$ & $95.4 \pm 2.0$ & $95.6 \pm 2.3$ & $\textbf{95.9 $\pm$ 1.4}$ \\ 
Small Bowel  & $72.2 \pm 16.2$ & $72.4 \pm 16.0$ & $75.4 \pm 16.0$ & $76.1 \pm 15.1$ & $\textbf{76.5 $\pm$ 15.3}$ \\ 
Pancreas  & $79.8 \pm 9.1$ & $79.9 \pm 10.6$ & $81.8 \pm 9.4$ & $81.8 \pm 8.5$ & $\textbf{82.1 $\pm$ 9.1}$ \\ 
Kidney L  & $95.7 \pm 1.2$ & $95.7 \pm 1.8$ & $95.8 \pm 1.4$ & $\textbf{96.0 $\pm$ 1.4}$ & $96.0 \pm 1.5$ \\ 
Kidney R  & $95.3 \pm 3.0$ & $95.5 \pm 2.8$ & $95.6 \pm 3.3$ & $95.6 \pm 2.7$ & $\textbf{95.7 $\pm$ 2.5}$ \\ 
Stomach  & $84.2 \pm 16.7$ & $85.0 \pm 15.8$ & $86.1 \pm 15.6$ & $86.8 \pm 13.6$ & $\textbf{87.5 $\pm$ 14.3}$ \\ 
Gallbladder  & $78.6 \pm 19.5$ & $78.2 \pm 20.1$ & $81.4 \pm 18.1$ & $\textbf{82.7 $\pm$ 17.2}$ & $82.2 \pm 17.7$ \\ 
\hline 
Average & $84.6$& $85.0$& $86.4$& $86.8$& $\textbf{86.9}$\\ \hline
\end{tabular}
&
\begin{tabular}{l l l l l l l}
\hline
Anatomy & U-Net & CA & C-CA & SCAA & SCAA$^*$ \\ \hline
Large Bowel  & $9.5 \pm 7.8$ & $9.7 \pm 8.3$ & $8.9 \pm 8.7$ & $7.1 \pm 4.6$ & $\textbf{6.6 $\pm$ 5.0}$ \\ 
Duodenum  & $7.8 \pm 4.9$ & $7.4 \pm 4.9$ & $6.6 \pm 4.7$ & $6.2 \pm 4.8$ & $\textbf{5.7 $\pm$ 4.1}$ \\ 
Spinal Cord  & $1.8 \pm 2.6$ & $1.9 \pm 2.8$ & $1.8 \pm 2.5$ & $1.9 \pm 3.0$ & $\textbf{1.6 $\pm$ 2.3}$ \\ 
Liver  & $3.9 \pm 3.9$ & $2.5 \pm 2.6$ & $2.6 \pm 3.4$ & $2.1 \pm 1.5$ & $\textbf{1.9 $\pm$ 1.4}$ \\ 
Spleen  & $6.5 \pm 12.8$ & $2.4 \pm 4.7$ & $2.5 \pm 7.3$ & $1.7 \pm 4.6$ & $\textbf{1.2 $\pm$ 0.7}$ \\ 
Small Bowel  & $7.8 \pm 7.3$ & $8.1 \pm 7.8$ & $9.0 \pm 11.4$ & $8.3 \pm 8.4$ & $\textbf{7.4 $\pm$ 7.1}$ \\ 
Pancreas  & $4.1 \pm 3.3$ & $3.9 \pm 3.9$ & $3.6 \pm 3.4$ & $3.5 \pm 3.5$ & $\textbf{3.3 $\pm$ 3.7}$ \\ 
Kidney L  & $1.5 \pm 1.4$ & $1.5 \pm 1.1$ & $1.9 \pm 5.4$ & $1.2 \pm 0.6$ & $\textbf{1.2 $\pm$ 0.4}$ \\ 
Kidney R  & $1.8 \pm 3.2$ & $1.5 \pm 1.6$ & $\textbf{1.3 $\pm$ 1.0}$ & $1.7 \pm 2.4$ & $1.3 \pm 1.1$ \\ 
Stomach  & $7.2 \pm 8.1$ & $6.6 \pm 7.4$ & $\textbf{5.7 $\pm$ 7.4}$ & $8.2 \pm 10.5$ & $5.9 \pm 7.9$ \\ 
Gallbladder  & $7.0 \pm 11.5$ & $6.0 \pm 7.6$ & $6.0 \pm 11.4$ & $4.9 \pm 9.1$ & $\textbf{3.1 $\pm$ 4.6}$ \\ 
\hline 
Average & $5.4$& $4.7$& $4.5$& $4.3$& $\textbf{3.6}$\\ \hline

\end{tabular}
\end{tabular}

\caption{\textbf{Ablation study on different ways of fusing 2D and 3D features.} \textbf{Left:} DSC (unit: \%). Higher the better. \textbf{Right:} 95\%HD (unit: mm). Lower the better. Bold numbers represent the best performance. CA stands for context-aware model, which does not progressively integrate 3D features from the 3D model. C-CA for center context-aware model, which only integrates the corresponding center slice from the 3D feature maps. SCAA for spatial context-aware self-attention model, which uses the MSFA module to aggregate 3D features from the whole 3D volume. SCAA$^*$ for model without concatenating $F_{globe}$ to the last feature map.
}
\label{table:ablation}
\end{scriptsize}
\end{threeparttable}
\end{table*}
\setlength{\tabcolsep}{6pt}

To compare different ways of integrating 3D features and demonstrate the contribution of each of the add-on modules in the proposed model, we conducted ablation studies with the following different settings: 1) model with only a 3D global descriptor $F_{globe}$ concatenated to the last feature map (CA), to demonstrate the importance of including 3D features progressively during feature extraction. 2) model without MSFA module, but only uses the corresponding center slice feature from the 3D context feature maps (C-CA), to show the effectiveness and importance of using self-attention to aggregate 3D features. 3) SCAA model without concatenating $F_{globe}$ to the last feature map (SCAA$^*$), to show whether $F_{globe}$ is crucial to improve the segmentation accuracy.

As we can see from \Cref{table:ablation}, by adding the 3D global descriptor to the 2D U-Net, CA outperforms the 2D U-Net by 0.4\% and lowers the 95\%HD by 0.7 mm, showing the importance of integrating the 3D holistic information. Next, by progressively integrating 3D features to 2D feature extractor, C-CA outperforms CA by 1.4\% in DSC and 0.2 mm in 95\%HD, showing the importance of integrating 3D context features hierarchically. Finally, by adding the MSFA module based on self-attention, SCAA outperforms C-CA by 0.4\%, demonstrating the effectiveness of the implemented self-attention mechanism to weigh the information from different adjacent slices. To find out which part of the 3D features contribute the most, we then compare the performance of SCAA and SCAA$^*$. SCAA$^*$ achieves average DSC of 86.9\% and average 95\%HD 3.6 mm,  while SCAA achieves 86.8\% and 4.3 mm. They are very close in terms of DSC and SCAA$^*$ wins on 95\%HD. This demonstrates that $F_{globe}$ has very little contribution, and progressively integrating 3D features with 2D encoder achieves the most performance gain.

For all following comparisons with other methods, we use the best performing configuration SCAA$^*$.

\subsection{Comparison with previous methods on ABD-110}
\begin{table*}
\centering
 \begin{threeparttable}
\begin{small}
\begin{tabular}{l l l l l l l l l}
\hline
Organ & U-Net$^1$ & Att. U-Net$^1$ & Att. U-Net & U-Net (3D patch)$^2$ & U\textsubscript{a}-Net & nnU-Net & SCAA \\ \hline
Large Bowel  & $80.5 \pm 9.4$ & $80.3 \pm 9.3$ & $80.2 \pm 8.7$ & $80.7 \pm 9.7$ & $77.1 \pm 10.4$ & $82.1 \pm 8.5$ & $\textbf{82.5 $\pm$ 9.2}$ \\ 
Duodenum  & $63.4 \pm 18.6$ & $64.5 \pm 18.1$ & $67.1 \pm 16.0$ & $70.2 \pm 16.3$ & $62.6 \pm 17.7$ & $\textbf{71.3 $\pm$ 15.8}$ & $70.7 \pm 17.5$ \\ 
Spinal Cord  & $90.3 \pm 3.8$ & $90.9 \pm 4.0$ & $89.8 \pm 3.9$ & $88.4 \pm 4.6$ & $\textbf{91.6 $\pm$ 3.5}$ & $89.5 \pm 4.6$ & $90.8 \pm 3.5$ \\ 
Liver  & $95.5 \pm 1.9$ & $95.7 \pm 1.5$ & $96.0 \pm 1.1$ & $95.9 \pm 1.2$ & $94.7 \pm 1.9$ & $96.4 \pm 1.0$ & $\textbf{96.4 $\pm$ 1.2}$ \\ 
Spleen  & $94.6 \pm 3.1$ & $95.1 \pm 2.8$ & $95.0 \pm 1.9$ & $92.9 \pm 13.2$ & $94.7 \pm 2.1$ & $93.8 \pm 12.7$ & $\textbf{95.9 $\pm$ 1.4}$ \\ 
Small Bowel  & $72.2 \pm 16.2$ & $73.7 \pm 16.6$ & $75.0 \pm 13.8$ & $73.7 \pm 15.6$ & $75.0 \pm 13.8$ & $75.7 \pm 14.3$ & $\textbf{76.5 $\pm$ 15.3}$ \\ 
Pancreas  & $79.8 \pm 9.1$ & $80.4 \pm 9.8$ & $79.0 \pm 10.0$ & $80.2 \pm 12.1$ & $76.3 \pm 11.6$ & $81.8 \pm 11.5$ & $\textbf{82.1 $\pm$ 9.1}$ \\ 
Kidney L  & $95.7 \pm 1.2$ & $95.6 \pm 1.6$ & $94.5 \pm 9.0$ & $94.2 \pm 9.0$ & $95.2 \pm 1.5$ & $94.8 \pm 9.1$ & $\textbf{96.0 $\pm$ 1.5}$ \\ 
Kidney R  & $95.3 \pm 3.0$ & $95.5 \pm 2.9$ & $94.9 \pm 3.1$ & $94.2 \pm 9.2$ & $94.8 \pm 2.5$ & $94.5 \pm 9.2$ & $\textbf{95.7 $\pm$ 2.5}$ \\ 
Stomach  & $84.2 \pm 16.7$ & $84.0 \pm 15.2$ & $85.4 \pm 15.6$ & $87.3 \pm 14.5$ & $83.1 \pm 14.1$ & $\textbf{88.0 $\pm$ 16.0}$ & $87.5 \pm 14.3$ \\ 
Gallbladder  & $78.6 \pm 19.5$ & $77.1 \pm 20.5$ & $78.2 \pm 19.1$ & $74.2 \pm 29.6$ & $72.6 \pm 26.8$ & $75.0 \pm 29.7$ & $\textbf{82.2 $\pm$ 17.7}$ \\ 
\hline 
Average & $84.6$& $84.8$& $85.0$& $84.7$& $83.4$& $85.7$& $\textbf{86.9}$\\ \hline
\end{tabular}
\end{small}

\begin{tablenotes}
   \item[1] 2D U-Net/Attention U-Net with adjacent 3 slices stacked into channel as input (2.5D). 
   \item[2] 3D U-Net with patch-wise input. 
\end{tablenotes}
\label{table:abd-110}

\caption{\textbf{Dice similarity coefficient (DSC) comparison on ABD-110.} Bold number means the best DSC performance. SCAA (proposed) achieves the highest DSC compared to other methods.
}
 \end{threeparttable}
\end{table*}

To compare with previous methods of multi-organ segmentation on the ABD-110 dataset, we ran the following representative algorithms: U-Net \cite{ronneberger2015u}, Attention U-Net \cite{oktay2018attention}, U\textsubscript{a}-Net \cite{tang2019clinically}, and nnU-Net \cite{isensee2018nnu}. U-Net is a well-established medical image segmentation baseline algorithm. Attention U-Net is a multi-organ segmentation framework that uses gated attention to filter out irrelevant response in the feature maps. U\textsubscript{a}-Net is a state-of-the-art end-to-end two-stage framework for multi-organ segmentation in the head and neck region. nnU-Net is a self-adaptive medical image semantic segmentation framework that wins the first in the Medical Segmentation Decathlon (MSD) challenge \cite{simpson2019large}. nnU-Net mainly consists of three main deep learning-based segmentation methods: a 2D U-Net (slice-wise), a 3D U-Net (patch-wise) and a coarse-to-fine cascade framework consisting of two 3D U-Nets. Its final model is an ensemble of the three methods. The above-mentioned works cover a wide range of algorithms for multi-organ segmentation and should provide a comprehensive and fair comparison to our proposed method on the in-house dataset. For 3D Attention U-Net, we followed the same preprocessing as in its original paper, to downsample the image to isotropic 2mm resolution due to GPU memory limitation. However, for all other methods, we feed the original CT image with its original image spacing.

The results are shown in \cref{table:abd-110}. First, by comparing 2D and 3D methods, we can see that the performance of 2D methods is on par with 3D methods on kidneys, spinal cord and liver, which is likely because those organs are usually large and have regular shapes. However, for organs like stomach, small and large bowels, 3D methods generally perform better. This may be because those organs often have more anatomical variance, and a 3D holistic understanding of the context is beneficial. Next, U\textsubscript{a}-Net was 1.4\% lower than 2D U-Net and 3.5\% lower than SCAA. This may be because U\textsubscript{a}-Net was designed mainly for the head and neck region where organs are relatively small and do not overlap too much with each other. The abdomen region, on the other hand, is more complicated as the bounding boxes of some organs overlap a lot with each other (e.g. large bowel and small bowel), which makes U\textsubscript{a}-Net less effective. Finally, comparing SCAA to nnU-Net, we find SCAA outperform nnU-Net by 1.2\%. The best configuration of nnU-Net on ABD-110 was ensemble of a 2D U-Net (slice by slice) and a 3D U-Net (patch-wise). Both nnU-Net and SCAA consider the fusion of 2D model and 3D model, but they implement it in different ways - nnU-Net uses ensemble to combine 2D and 3D models while SCAA integrates 3D model into 2D model in an end-to-end fashion and jointly optimizes both models. This improvement then is likely due to the soft attention mechanism that allows SCAA to filter and extract relevant features from the large 3D context and better fuse the 2D and 3D models. Altogether, we demonstrated the effectiveness of the proposed method, which achieves an average DSC of 86.9\% on the in-house dataset.

\subsection{Performance on MICCAI2015}
\setlength{\tabcolsep}{2pt}
\begin{table*}
\centering
\begin{small}
\begin{tabular}{@{\extracolsep{0pt}}l c c c c c c c c c c c c c c c c c c@{}}
\hline
\multirow{2}{*}{Study} & Brain Stem & Mandible & Optic Chiasm & \multicolumn{2}{c}{Optic Nerve} & \multicolumn{2}{c}{Parotid} & \multicolumn{2}{c}{SMG} & \multirow{2}{*}{Avg.} \\ \cline{5-6} \cline{7-8} \cline{9-10}
 & & & & L & R & L& R & L & R & \\ \hline 
Raudashl \etal \cite{raudaschl2017evaluation} & 88.0 & 93.0 & 55.0 & 62.0 & 62.0 & 84.0 & 84.0 & 78.0 & 78.0 & 76.0\\ 
Fritscher \etal \cite{fritscher2016deep} & & & 49.0 $\pm$ 9.0& & & 81.0 $\pm$ 4.0& 81.0 $\pm$ 4.0& 65.0 $\pm$ 8.0& 65.0 $\pm$ 8.0& -\\ 
Ren \etal \cite{ren2018interleaved} & & & 58.0 $\pm$ 17.0& 72.0 $\pm$ 8.0& 70.0 $\pm$ 9.0& & & & & -\\ 
Wang \etal \cite{wang2018hierarchical} & \textbf{90.0 $\pm$ 4.0}& 94.0 $\pm$ 1.0& & & & 83.0 $\pm$ 6.0& 83.0 $\pm$ 6.0& & & -\\ 
Zhu \etal \cite{zhu2018anatomynet} & 86.7 $\pm$ 2.0& 92.5 $\pm$ 2.0& 53.2 $\pm$ 15.0& 72.1 $\pm$ 6.0& 70.6 $\pm$ 10.0& 88.1 $\pm$ 2.0& 87.3 $\pm$ 4.0& 81.4 $\pm$ 4.0& 81.3 $\pm$ 4.0& 79.2\\ 
Tong \etal \cite{tong2018fully} & 87.0 $\pm$ 3.0& 93.7 $\pm$ 1.2& 58.4 $\pm$ 10.3& 65.3 $\pm$ 5.8& 68.9 $\pm$ 4.7& 83.5 $\pm$ 2.3& 83.2 $\pm$ 1.4& 75.5 $\pm$ 6.5& 81.3 $\pm$ 6.5& 77.4\\ 
Nikolov \etal \cite{nikolov2018deep} & 79.5 $\pm$ 7.8& 94.0 $\pm$ 2.0& & 71.6 $\pm$ 5.8& 69.7 $\pm$ 7.1& 86.7 $\pm$ 2.8& 85.3 $\pm$ 6.2& 76.0 $\pm$ 8.9& 77.9 $\pm$ 7.4& -\\ 
Tang \etal \cite{tang2019clinically} & 87.5 $\pm$ 2.5& 95.0 $\pm$ 0.8& 61.5 $\pm$ 10.2& 74.8 $\pm$ 7.1& 72.3 $\pm$ 5.9& 88.7 $\pm$ 1.9& 87.5 $\pm$ 5.0& 82.3 $\pm$ 5.2& 81.5 $\pm$ 4.5& 81.2\\ 
Guo \etal \cite{guo2020organ} & 87.6 $\pm$ 2.8& 95.1 $\pm$ 1.1& \textbf{64.5 $\pm$ 8.8}& 75.3 $\pm$ 7.1& 74.6 $\pm$ 5.2& 88.2 $\pm$ 3.2& 88.2 $\pm$ 5.2& \textbf{84.2 $\pm$ 7.3}& \textbf{83.8 $\pm$ 6.9}& 82.4\\ \hline
SCAA (proposed) & 89.2 $\pm$ 2.6& \textbf{95.2 $\pm$ 1.3}& 62.0 $\pm$ 16.9& \textbf{78.4 $\pm$ 6.1}& \textbf{76.0 $\pm$ 7.5}& \textbf{89.3 $\pm$ 1.5}& \textbf{89.2 $\pm$ 2.3}& 83.2 $\pm$ 4.9& 80.7 $\pm$ 5.2& \textbf{82.6}\\ 
\hline
\end{tabular}
\caption{\textbf{Comparison of DSC with previous methods on the MICCAI 2015 9 organs segmentation challenge.} Numbers are the higher the better (best in bold).}
\label{table:miccai15}
\end{small}
\end{table*}
A second multi-organ segmentation dataset from MICCAI 2015 organ-at-risk (OAR) segmentation challenge\cite{raudaschl2017evaluation} was used for evaluation. First, as we can see from \Cref{table:miccai15}, SCAA outperforms \cite{nikolov2018deep} by 4.2\%. \cite{nikolov2018deep} used a combination of 3D and 2D convolution on 21 stacked slices for OAR segmentation. This shows the use of larger context information is beneficial for a good segmentation accuracy. Next, by comparing SCAA to AnatomyNet \cite{zhu2018anatomynet} which is a 3D model that takes the whole-volume CT as input, SCAA was 2.4\% higher. This is likely due to the attention mechanism that helps the model to filter irrelevant features from the entire volume. Also, SCAA outperforms U\textsubscript{a}-Net \cite{tang2019clinically} by 1.4\%. U\textsubscript{a}-Net is an end-to-end two-stage model that first detects bounding box of OARs and then segments organs within the bounding box. SCAA performed better, partly because SCAA did not enforce a 'hard' attention (bounding box) but rather use 'soft' attention to enable the model focus on a smaller region. This keeps SCAA away from potential bounding box regression error and missing detection. Finally, SCAA outperforms previous state-of-the-art method \cite{guo2020organ} in 5 out of 9 organs and achieves an average DSC of 82.6\%, 0.2\% higher than the state-of-the-art method. Also note that \cite{guo2020organ} is a two-stage segmentation framework, which consists of two DCNNs. Our proposed method (SCAA), however, is a one-stage end-to-end solution for multi-organ segmentation, requiring less training time and computation, as well as fewer parameters.

\subsection{Memory consumption}
\begin{table*}
\centering
 \begin{threeparttable}
\begin{tabular}{l c c c l}
\hline
\textbf{Method} & \textbf{Batch size} &  \textbf{Estimate (GB)} & \textbf{Actual (GB)} & \textbf{\# of parameters} \\ 
\hline
2D U-Net \cite{ronneberger2015u} &  4 & 2.86 & 3.35 & 34.51 M\\
3D U-Net \cite{oktay2018attention} & 1 & 27.96 & out of memory & 5.88 M\\
3D Attention U-Net \cite{oktay2018attention} & 1 & 17.31 & out of memory & 6.40 M\\
\hline
SCAA (3D encoder) & 1 & 3.22 & - & - \\
SCAA (2D U-Net \& MSFA) & 4 & 2.13 & - & -\\
SCAA (total) & & 5.35 & 6.44 & 7.82 M\\
\hline
\end{tabular}

  
\caption{\textbf{GPU memory consumption comparison using whole-volume image as input of, and number of parameters for different methods on ABD-110.} We used PyTorch as the deep learning framework to measure the actual GPU memory cost.}
\label{table:gpu-memory}
 \end{threeparttable}
\end{table*}

One advantage of the proposed method is that it significantly reduces the GPU memory while at the same time preserves the large 3D context features. 
To demonstrate GPU memory consumption when using whole-volume as input, we estimated and measured the actual GPU memory cost (using PyTorch as framework) for different 3D models during training in \Cref{table:gpu-memory}. We made several assumptions: \romannumeral1) the input image volume is of size 256 $\times$ 256 $\times$ 256. This is the size used for whole-volume input with original image spacing. \romannumeral2) we only take the memory cost of storing tensor and its gradient after each convolution and batch/instance normalization layer into consideration, because they consume the most GPU memory. \romannumeral3) each number is a floating point number (32 bits). For 2D U-Net, the numbers of channel for the five scales are 64, 128, 256, 512 and 1024 respectively, as in the original implementation \cite{ronneberger2015u}. For the 3D U-Net, as the network has more parameters in convolution kernels, fewer channels are used in practice (16, 32, 64, 128 and 256) \cite{oktay2018attention}. The memory cost and number of parameters of \cite{ronneberger2015u} and \cite{oktay2018attention} were computed by running their released code. 
The GPU memory cost is for the training phase, and that of inference is approximately half of the values in \Cref{table:gpu-memory}. The actual cost is computed by running the algorithm on a GTX 1080 Ti GPU card (12 GB memory).

As seen from \Cref{table:gpu-memory}, compared to most 3D U-Net based methods, we only require 6.44 GB total memory for a batch size of four during training, which is approximately 35.1\% of the 3D Attention U-Net, demonstrating the efficiency of the proposed method. Moreover, our method supports distributing batches among multiple GPU devices, which is more scalable than previous 3D methods. 

\subsection{Visualization}

\begin{figure}
\centering
  \includegraphics[width=0.9\linewidth]{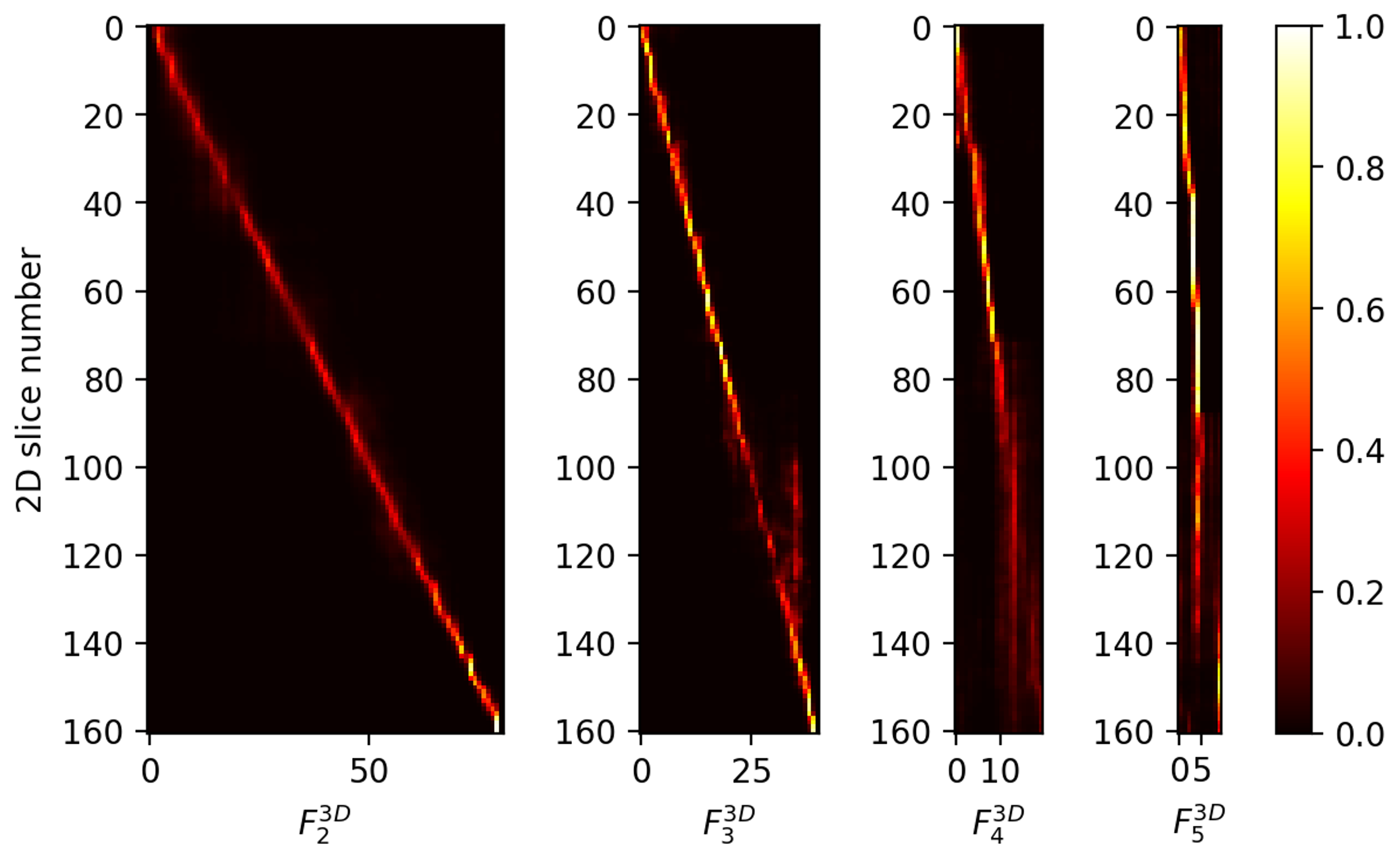}
\caption{Attention vector learnt by the proposed method. Y-axis is the slice number of the CT image, and the X-axis is the slice number (depth) of the 3D feature map $F_i^{2D}$.}
\label{fig:attention}
\end{figure}

\begin{figure*}
\centering
  \includegraphics[width=0.8\linewidth]{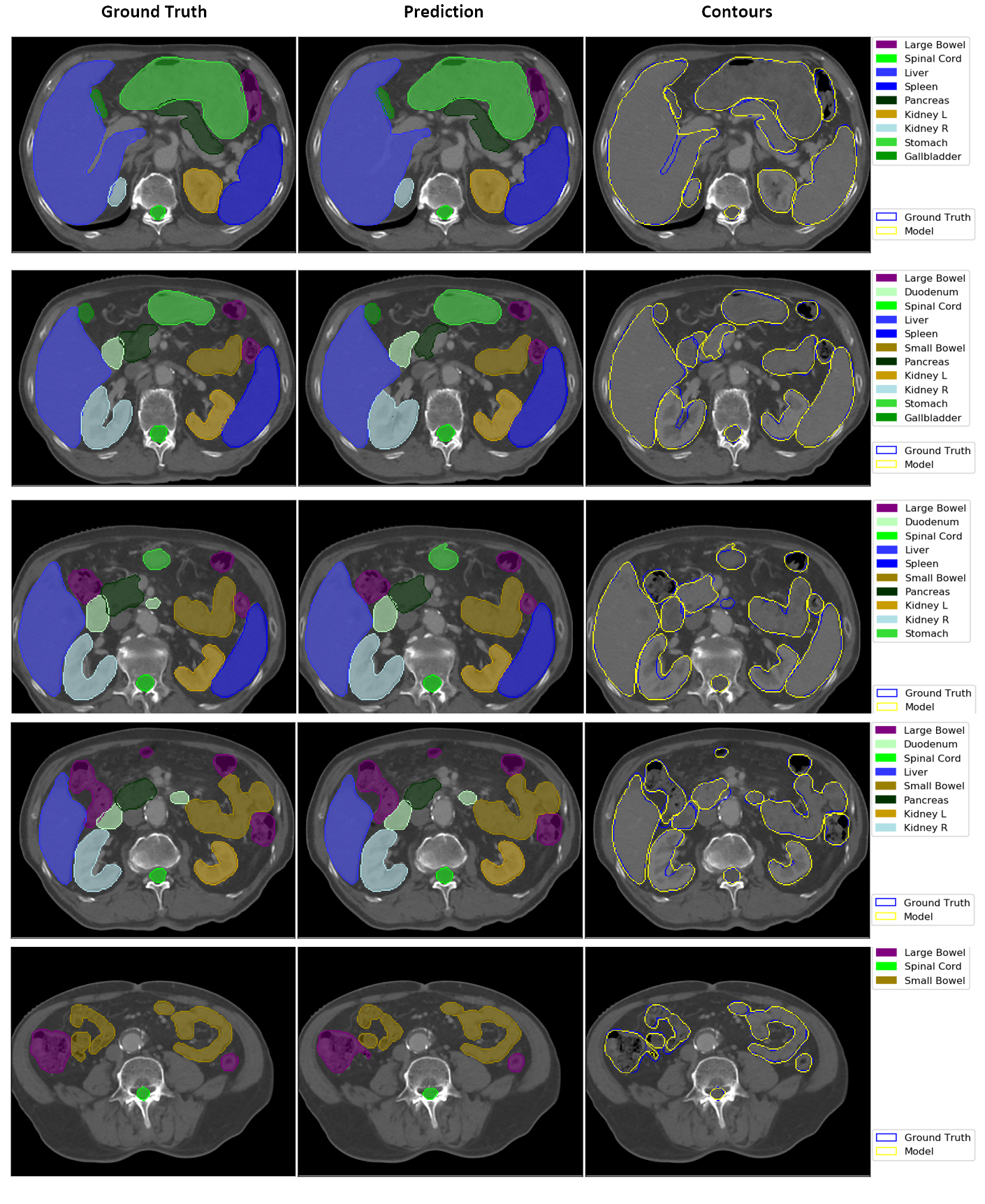}
\caption{\textbf{A CT image from ABD-110 dataset}. The first and second columns are the ground truth and predicted mask overlayed on the original CT image slice, respectively. The third column shows the comparison of contours of the ground truth and predicted mask on the same slice.}
\label{fig:visualization}
\end{figure*}

We visualize the attention vector $\mathbf{a_i}$ learnt for $F_i^{3D}$ \Cref{fig:attention} and the prediction of a random CT image from the ABD-110 dataset \Cref{fig:visualization}. As seen from \Cref{fig:attention}, the 3D slice features that are useful when segmenting each 2D slice are mostly its adjacent slices. This accords with the intuition that the most prominent and useful 3D information should be mostly from its neighbouring slices. But it is also important to incorporate full 3D context information progressively. We have demonstrated the effectiveness of the self-attention mechanism in \Cref{subsection:ablation} by comparing to C-CA that only integrates the corresponding center slice feature from the 3D feature map.



\section{Conclusion}
In this paper, we propose a Spatial Context-aware Self-Attention model for multi-organ segmentation. The proposed model uses a self-attention mechanism to filter useful 3D contextual information from the large 3D whole-volume CT image to guide the segmentation of 2D slice. It addresses the GPU memory concerns that common whole volume-based 3D methods confront. Experiments on two multi-organ segmentation datasets demonstrate the state-of-the-art performance of the proposed model.

\newpage

{\small
\bibliographystyle{ieee_fullname}
\bibliography{egbib}
}

\end{document}